\documentclass[]{spie}  

\usepackage{amsmath,amsfonts,amssymb}
\usepackage{graphicx}
\usepackage{siunitx}
\usepackage{aas_macros}
\usepackage[colorlinks=true, allcolors=blue]{hyperref}

\title{PROTOCALC: an artificial calibrator source for CMB telescopes}

\author[a,b]{Gabriele Coppi}
\author[a]{Giulia Conenna}
\author[a]{Sofia Savorgnano}
\author[c]{Felipe Carrero}
\author[c]{Rolando D\"{u}nner-Planella}
\author[c,d]{Nicholas Galitzki}
\author[a,b]{Federico Nati}
\author[a,b]{Mario Zannoni}
\affil[a]{Department of Physics, University of Milano-Bicocca, Piazza della Scienza 3, 20126 Milano, Italy}
\affil[b]{National Institute for Nuclear Physics (INFN), Sezione di Milano-Bicocca, Piazza della Scienza 3, 20126 Milano, Italy}
\affil[c]{Instituto de Astrof\'{i}sica and Centro de Astro-Ingenier\'{i}a, Facultad de F\'{i}sica, Pontificia Universidad Cat\'{o}lica de Chile, Av. Vicu\~{n}a Mackenna 4860, 7820436, Macul, Santiago, Chile}
\affil[d]{Department of Physics, University of California San Diego, La Jolla, CA 92093, USA}

\authorinfo{Further author information: (Send correspondence to Gabriele Coppi)\\E-mail: gabriele.coppi@unimib.it}

\begin{document} 
\maketitle

\begin{abstract}
Cosmic Microwave Background experiments need to measure polarization properties of the incoming radiation very accurately to achieve their scientific goals. As a result of that, it is necessary to properly characterize these instruments. However, there are not natural sources that can be used for this purpose. For this reason, we developed the \textbf{PROTOtype CALibrator for Cosmology}, PROTOCALC, which is a calibrator source designed for the \SI{90}{\giga\hertz} band of these telescopes. This source is purely polarized and the direction of the polarization vector is known with an accuracy better than \SI{0.1}{\degree}. This source flew for the first time in May 2022 showing promising result.
\end{abstract}


\section{INTRODUCTION}
\label{sec:intro}

Cosmic Microwave Background (CMB) polarization instruments need a very accurate control on the systematic errors introduced by absolute polarization orientation and polarized beam patterns, which limit the accuracy on multiple astrophysical signals such as the Inflationary Gravitational Waves and influences the reconstruction of the gravitational lensing effects on the CMB. Besides, an absolute calibration of the polarization angle of the CMB photons would enable the detection of signatures of parity-violating mechanisms in the early Universe, such as Cosmic Birefringence. Estimates for next-generation experiments such as Simons Observatory show that the accuracy required on the absolute polarization angle for measuring the tensor-to-scalar ratio $r$ with a delta lower than $2\cdot10^{-4}$ is around \SI{0.2}{\degree}\cite{abitbol2021}. However, this estimate is based on EB nulling techniques that are based on the assumption of the absence of parity-violation mechanisms. At the moment there are some evidence of the presence of Cosmic Birefringence\cite{Minami2020}. For this reason, it is necessary to have an independent calibrator for CMB telescopes. Unfortunately the best natural polarization calibrator, TAU-A, is not known well enough for CMB appplication\cite{Aumont2020}. Given this background we present the design and the realization of an artificial calibrator which should provide the polarization angle with an accuracy better than \SI{0.1}{\degree}. This system has been thought especially for the telescopes on the Atacama desert, such as Simons Observatory\cite{SO2019} and CLASS\cite{Class2014}, but can easily adapted to other sites.

\section{PROTOCALC}
\label{sec:protocalc}

PROTOCALC (PROTOtype CALibrator for Cosmology) is a project funded as a Marie-Curie Fellowship under the Horizon-2020 Program. The goal of the project is to develop a \SI{90}{\giga\hertz} polarization calibrator for CMB Telescopes with a polarization angle accuracy of \SI{0.1}{\degree}. To achieve this result, we need to develop a calibrator that is flexible enough and can be used at multiple telescope sites around the Earth. Therefore, we decided to develop the concept of using a drone as a carrier for the calibrator\cite{Nati2017,Dunner2020}. In order to keep the project as simple as possible, we decided to use a commercially available drone like the DJI Matrice 600 Pro. Furthermore, given the project accuracy goal, we need to have the source as stable as possible, so it will be installed on a DJI RONIN MX gimbal to improve the stability during the flights. 

Finally, given the prototype nature of the project, we decided to design a source in a way that we can upgrade multiple components in the future to extend the frequency band or improve the components.

\begin{figure}
    \centering
    \begin{minipage}{0.45\textwidth}
        \centering
        \includegraphics[height=1\textwidth]{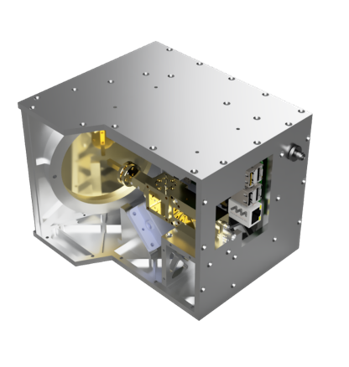}
        \caption{Rendering of the PROTOCALC payload.}
        \label{fig:render}
    \end{minipage}\hfill
    \begin{minipage}{0.45\textwidth}
        \centering
        \includegraphics[height=1\textwidth]{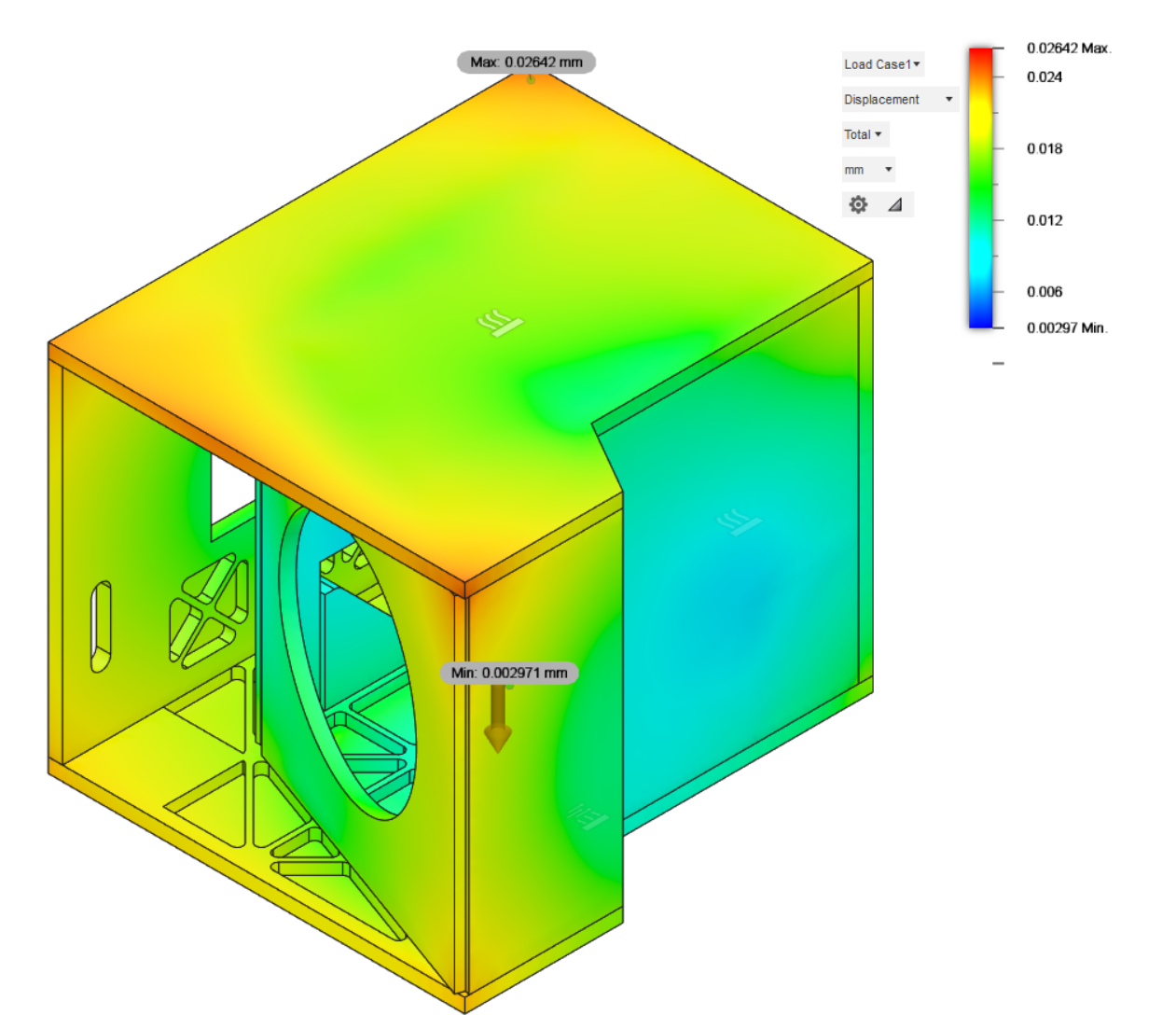}
        \caption{Displacement due to thermal contraction as simulated using Autodesk Fusion360.}
        \label{fig:displacement}
    \end{minipage}
\end{figure}

\section{DESIGN}
\label{sec:design}

Given that calibrator is effectively a payload on board a drone, we do not have direct control of the source. Consequentially, we want the payload to be as autonomous as possible and we require a flight computer that controls everything on board. For this reason, the core of the payload is a Raspberry Pi3 (RPi) computer. Moreover, a calibrator like this one has two particular requirements that we need to consider in designing the payload. The first one is the ability to deliver and maintain throughout the observations the required polarization angle accuracy of \SI{0.1}{\degree}. Instead, the second is mechanical. Indeed, we need to consider that the system needs to fit inside a volume that is given by the gimbal capacity and its weight needs to be within the drone specification at liftoff. 

\subsection{Pointing Subsystem}
\label{subsec:pointing}

To achieve the goal of \SI{0.1}{\degree}, we need to pay particular attention to the attitude and position of the calibrator. The DJI drone is already designed to use Real-Time Kinematic (RTK) to provide location with an accuracy of few \SI{}{\cm}. However, we wanted the source to be as independent as possible from the drone, therefore we use a UBLOX GPS receiver that can be used for RTK applications or for Post-Processing Kinematic (PPK) after the flight. Given the multiple outputs available on the UBlox board, we used the GPS information also as a timing server for the RPi to simplify the synchronization later with the data received by the telescopes. The position, and error, on the location of the drone are fundamental for the telescope pointing, but they do not provide any specific information regarding the polarization vector direction. This information can be extracted from the payload attitude and its yaw, pitch and roll angles. In particular, the most important one for the polarization angle is the roll angle, which defines how much the polarization vector is rotated. To measure this angle we use a system with an inclinometer and and a camera used for photogrammetry\cite{Dunner2020}. The camera used in the system is a commercially available Sony RX0-MII controlled by a RPi as explained in \ref{subsec:software}. To use photogrammetry for attitude reconstruction, we use a series of georeferenced target on the ground and then we analyze the photos taken by the camera to properly estimate the Euler angles of the payload.  

\subsection{Mechanical Design}
\label{subsec:mechanical_design}

The payload design needs to consider the weight limitations of the drone platform and the volume available in the gimbal mount. These constraints limit the weight to maximum \SI{2}{\kilo\gram} and a volume that can represented by a box of 16 x 16 x 13 cm$^3$. For thermal management reason, we decided to use Aluminum-6061 as the main material for the payload. Indeed, the choice of this material was due to its high thermal conductivity at room temperatures, so that we can avoid any kind of active coolers on high-power components such as the multiplier or the frequency synthesizer and use passive dissipation. We chose a multiple parts design to allow us easy access also on the field to verify each component in case of emergency. The installed inclinometer uses a \SI{45}{\degree} mount to guarantee that is almost flat during observation that are performed with the gimbal at approximately \SI{-45}{\degree}. One of the key element to guarantee an accurate measurements of the roll angle is the correct alignment between the photogrammetry camera and the polarization defining filter. While we calibrated the alignment in laboratory as explained in \ref{subsec:optical_calibration}, the use of a material like Al-6061 guarantees a higher accuracy during manufacturing compared to 3D printed plastics solutions, giving us better control of the position of the camera and the polarizing grid. A rendering of the design is presented in Fig.\ref{fig:render}.

\subsection{Mechanical and Thermal Simulation}
\label{subsec:simulations}

In order to asses the goodness of the design we perform multiple simulations. Given the environment where the system is going to be deployed the most important simulation is the thermo-mechanical one. Indeed, this guarantees that the choice of an aluminum payload for thermal reason is justified and that eventual thermal contractions between the camera and the filter (and the inclinometer) are negligible. We performed the simulation using Autodesk Fusion360 where we set as a mechanical loading only the gravity and as thermal load we included the various thermal sources like the RPi, Multiplier and Valon. The cooling is provided by radiation and convection, whose coefficient takes into account the fact that the payload operates at an altitude of $\sim$\SI{5200}{\meter} where there the atmospheric pressure is only half compared to the sea level. Moreover, we consider that the thermal stress free temperature is \SI{20}{C} like the laboratory, while the operational temperature is \SI{0}{C}. The results are presented in Fig.\ref{fig:displacement}. As it is possible to notice the relative displacement between the camera and the polarization grid mounting holes is negligible, $\le$ \SI{10}{\micro\meter}. 

\subsection{Software Design}
\label{subsec:software}

As mentioned before, the flight computer of PROTOCALC is RPi 3. We chose this model because it has multiple threads and its power consumption is around \SI{2.5}{\watt} during normal operation for our application (between 3 and 4 times lower than a RPi4). PROTOCALC is run by a python code called PORTER, \emph{Protocalc cOntRol sysTEm pRogram}. The software runs automatically as soon as the RPi is turned on. The first operation is checking the time and the GPS and update the clock of the RPi. The code is threaded, so that the data acquisition from the different sensors can be parallel. Both the GPS and the inclinometer have their own thread to read the data and save them on the RPi internal memory. Another thread takes care of the camera control. This control uses the Sony Remote Camera SDK written in C++. Since the metadata in the photos or videos from the Sony RX0MII has a precision of \SI{1}{\second}, to properly time stamping them we create a log file where we saved the time at which the command is sent (to take the photos or the start recording the video). The camera control can also set other parameters of the camera, however those are not set automatically but are read from a configuration file. The only component that is not currently threaded is the frequency synthesizer control. This is because at the moment we set the Valon-5019 at the beginning (just after the time correction) and it uses those parameters (frequency, power output and amplitude modulation) throughout the entire flight. 

\section{RF Configuration}
\label{sec:rf_configuration}

\begin{figure}
    \centering
    \includegraphics[width=\textwidth]{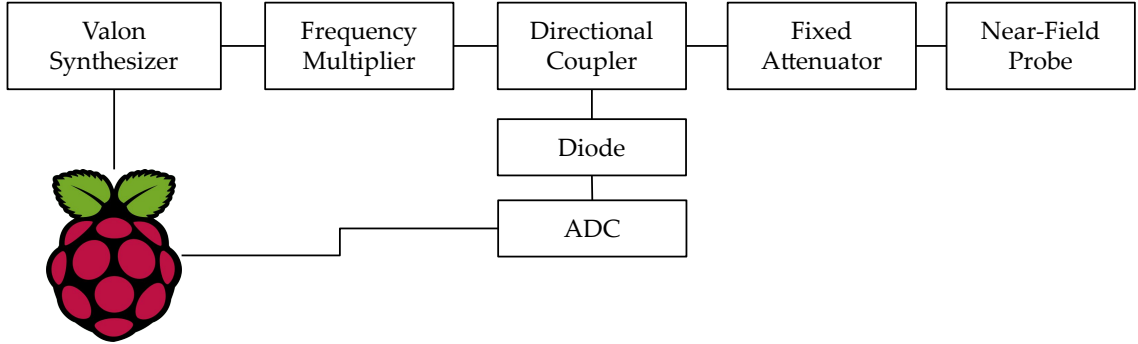}
    \caption{PROTOCALC RF chain schematic.}
    \label{fig:schematic}
\end{figure}

At the core of PROTOCALC we have the source that generates the signal that will be observed by the telescopes. As mentioned before, the output frequency of PROTOCALC is in the W-Band, so between \SI{75}{\giga\hertz} and \SI{115}{\giga\hertz}. The current RF configuration allows the output of a signal with a very small width ($\simeq$ \SI{10}{\kilo\hertz}) and we decided to have as our main output a signal at \SI{90}{\giga\hertz}. To generate this signal, we use a Valon-5019 frequency synthesizer that generates a signal at \SI{15}{\giga\hertz}. This signal is then multiplied by a 6 factor by the Erevant frequency multiplier at \SI{90}{\giga\hertz}. At the output of the multiplier, we have a directional coupler. At the output of the coupled port we installed a diode to control the output of the multiplier. The output of the transmitted port is attenuated using a passive \SI{30}{\decibel} attenuator and finally there is a near field probe which is the antenna output towards the telescope. The signal from the near-field probe is already polarized as in any waveguide, however a polarizing grid is put in front of the antenna to define the polarization angle and give the possibility to modify the angle simply rotating the filter. A full representation of the configuration is presented in Fig.\ref{fig:schematic}. As it is possible to notice in the schematic, everything is controlled by a RaspberryPi. This computer set the correct parameters for the the frequency and power output of the Valon-5019. Connected to RPi, there is also an ADC that converts into digital the signal output of the diode. The signal is recorded and time-tagged on the RPi on-board memory for postprocessing analysis. 

\section{LABORATORY CALIBRATION}

In order to properly understand the characteristics of PROTOCALC, we performed a series of in-lab tests to calibrate the instrument and study the behaviour of the different components. We divide the tests performed into two different categories: 

\begin{itemize}
    \item RF tests and calibration: to verify the proper behaviour of the RF chain
    \item Optical tests and calibration: to verify the beam output and the camera alignment
\end{itemize}

\subsection{Source Calibration}
\label{subsec:source_calibration}

To completely characterize the RF components we performed a series of tests on all the components of the RF chain. The passive components were analyzed using a vector network analyzer (VNA) to verify the compliance with the advertised characteristics. For the active components such as the Valon-5019 and the frequency multiplier we performed multiple tests using a signal/spectrum analyzer. 

In the the first phase of tests we studied the performance of the Valon-5019. In particular we tested the frequency output stability and accuracy of the synthesizer and the output power. We tested the Valon up to \SI{19}{\giga\hertz}\footnote{With the new firmware update the Valon-5019 can achieve a maximum of \SI{20}{\giga\hertz}, however this firmware was not available at the time of testing.} from \SI{10}{\giga\hertz}. The first test that we performed was the output frequency verification, in particular we wanted to test how the frequency generated differs from the nominal frequency that we set using the RPi. As you can notice in the Table \ref{tab:frequencyshift}, the difference is negligible and will not be detectable by any CMB telescope. 

\begin{table}
    \centering
	\begin{tabular}{c||c|c}
		$f$ nominal [GHz] & $f_{m}$ [GHz] &  $\sigma_{f_m}$ [GHz] \\
		\hline
		10.0 & 9.999997824 & 4e-09  \\
		11.0 & 10.999997627 & 2e-09 \\
		12.0 & 11.999997349 & 5e-09 \\
		13.0 & 12.999997163 & 5e-09 \\
		14.0 & 13.999996877 & 3e-09 \\
		15.0 & 14.999996731 & 4e-09 \\
		16.0 & 15.999996506 & 5e-09 \\
		17.0 & 16.999996418 & 9e-09 \\
	    18.0 & 17.99999607 & 5e-09  \\
		19.0 & 18.999995943 & 8e-09 \\
	\end{tabular}
	\caption{Table comparing the nominal frequency and the measured frequency of the Valon-5019.}
	\label{tab:frequencyshift}
\end{table}

\begin{figure}
    \centering
    \includegraphics[scale=0.4]{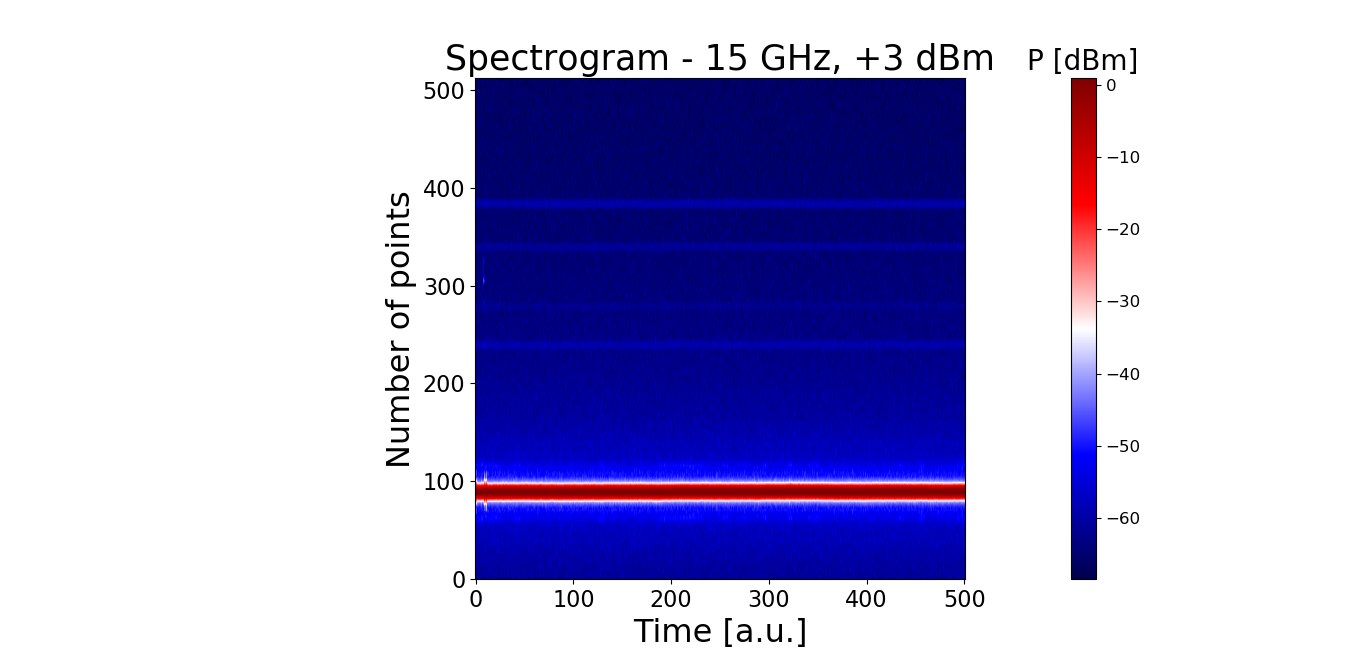}
    \caption{Spectrogram as acquired by the spectrum analyzer at \SI{15}{\giga\hertz}}
    \label{fig:spectrogram}
\end{figure}

To verify the frequency stability and the performance of the local oscillator inside the Valon-5019, we recorded with the signal analyzer its spectrogram output. From this spectrogram we analyzed how much the central frequency at each timestamp vary and we fitted a line to verify the shift. In Fig.\ref{fig:spectrogram}, it is possible to see the gradient of this line and the results show a negligible shift for the application of interest. 

\begin{figure}
    \centering
    \includegraphics[scale=0.5]{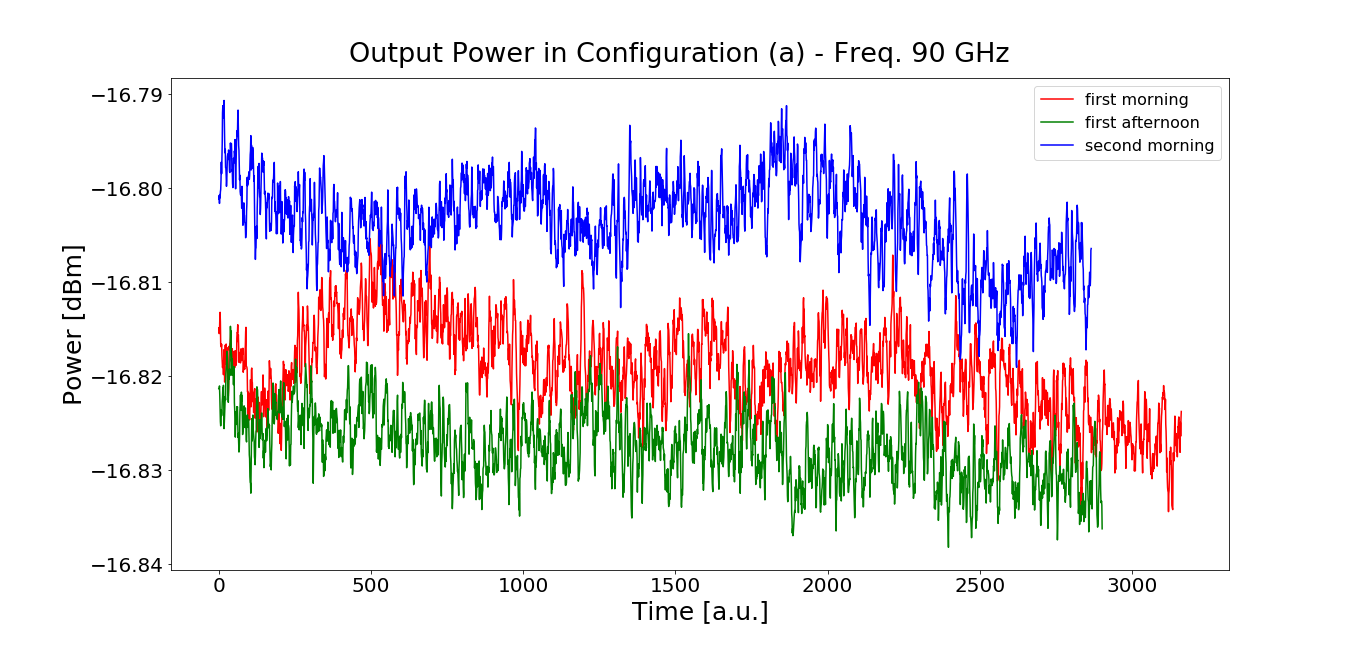}
    \caption{Stability of the power output in three different days. On the first and last day, the data was acquired in the morning while in the second day we acquired data in the afternoon.}
    \label{fig:stability}
\end{figure}

To further investigate the frequency stability, we measured the stability of the power output at \SI{90}{\giga\hertz} with the full RF chain installed in different days. As it is possible to notice in Fig.\ref{fig:stability}, the output change in different days can be neglected for the PROTOCALC application. 

Finally, we tested the output power compared to nominal output power that we set with the RPi. This test is particular crucial because the frequency multiplier requires an input power of \SI{3}{\decibel\meter} to properly function. From our measurements, we notice that the Valon-5019 performs accordingly to the power output specifications at all frequencies but \SI{12}{\giga\hertz} and \SI{13}{\giga\hertz}. This means that when we need to use those lower frequencies we need to send a command with a higher power value to compensate for this effect.

\subsection{Optical Calibration}
\label{subsec:optical_calibration}

\begin{figure}
    \centering
    \includegraphics[scale=0.5]{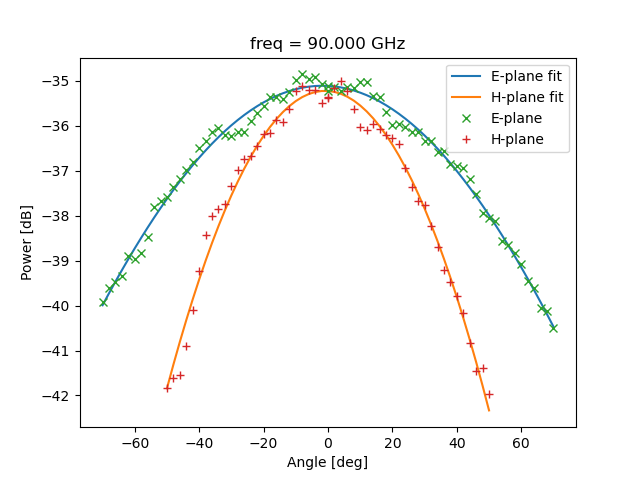}
    \caption{E-plane and H-plane measurements for the near field probe at \SI{90}{\giga\hertz}.}
    \label{fig:horn}
\end{figure}

For this part of the calibration work, we focused our attention to the near field probe and to measure the misalignment between the photogrammetry camera and the polarizing grid. To characterize the near field probe, we use a VNA and measure the E and H planes. The resulting data are fitted and then compared to data provided by the manufacturer. As can been seen in the Fig. \ref{fig:horn}, our measurements confirm the expected values.

\begin{figure}
    \centering
    \includegraphics[width=0.7\textwidth]{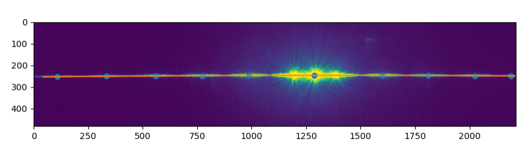}
    \caption{Fitting of the diffraction pattern. The blue dots are the centroid of each maxima and the orange line is the fitted parabola. }
    \label{fig:alignement}
\end{figure}

While this test was just done to confirm the expected value, the alignment between the camera and the polarizing grid is a calibration specific of the instrument. To perform this measurement, we applied a similar methodology than the one developed in\cite{Carrero2021}. We shine a laser through the polarization grid and we shoot a series of photos with the photogrammetry camera and we analyze each photo to estimate the relative rotation. To fully analyze each photo, we first correct for the camera distortions\footnote{The camera was previously calibrated using a chessboard like image and openCV routines.}. At this point, we preprocess the image using a gaussian de-noise filter and we use a threshold to distinguish the background from the diffraction maxima in the image. To recognize each maximum separately, we use the scikit-image package and look for the contours of each one using a contrast technique. Due to noise or other imperfection in the camera sensor, it may happen to have a cold spot in some of the maxima. To avoid that our region finding algorithm is affected by the presence of these spots, we still use scikit-image to fill these gaps. Once the regions are correctly identified, we need to analyze the image to estimate the angle. To perform this task, we developed a different fitting technique with respect to what has been done in\cite{Carrero2021}. Similarly to this work, we find the presence of second order distortions that do not allow to use a simple linear fit, so we need to use a parabola. However, since we need to measure how much this parabola is rotated we fit a generic conic where we imposed the parabola condition and from there extracting the rotation. This analysis is performed on the image in two different ways. In the first method, we use only the centroid of each maximum while in the second method we compute a centroid for each pixel column, giving us significantly more points. We did this for multiple images and then we compute the rotation angle using an average and the error is its standard deviation. An example of a fit is present in Fig. \ref{fig:alignement}. We did these analysis on multiple acquisitions and the results are presented in table \ref{tab:offset}. For each dataset we took 40 images and we analyze only 7 random photos from the set. Repeating the analysis with a different subsample does not result in any significant deviation from the results reported here. We need to note that this calibration needs to be performed each time that the filter and/or the camera are reinstalled, since each time we may have a slightly different rotation angle. 

\begin{table}[]
    \centering
    \begin{tabular}{c|cc}
         Dataset & $\psi$ (\SI{}{\degree}) & $\Delta\psi$ (\SI{}{\degree})  \\
         \hline
         \#1 & 0.164 & 0.007 \\
         \#2 & 0.167 & 0.011 \\
         \#3 & 0.164 & 0.003 
    \end{tabular}
    \caption{Angle offset between the camera and the polarization grid and its uncertainty.}
    \label{tab:offset}
\end{table}

\section{FIRST FLIGHT}
\label{sec:first_flight}

\begin{figure}
    \centering
    \begin{minipage}{0.3\textwidth}
        \centering
        \includegraphics[scale=0.6]{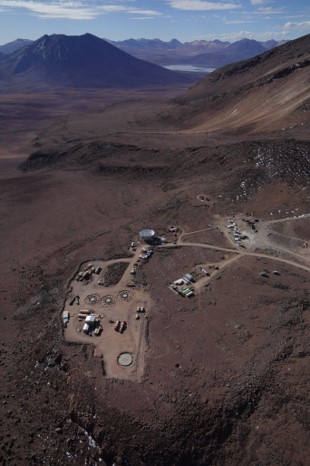}
        \caption{Shot taken by the photogrammetery camera during one flight. It is possible to notice some small with dots that represents the photogrammetry targets.}
        \label{fig:view}
    \end{minipage}\hfill
    \begin{minipage}{0.55\textwidth}
        \centering
        \includegraphics[scale=1]{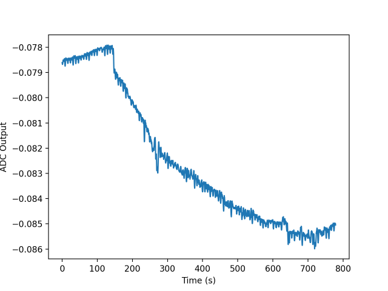}
        \caption{RAW value as measured during the flight. The change of regime from almost constant to transient happens when the drone start the scanning.}
        \label{fig:adc}
    \end{minipage}
\end{figure}

Between the end of April 2022 and the beginning of May 2022, we performed our first flight above the Cerro Toco Plateau in Chile. We did several tests with multiple telescopes (CLASS, Polarbear2 and ACT) observing the source. While the analysis of the signal is still ongoing it is possible to share some information about the drone flying. The drone was flown at approximately \SI{350}{\meter} above the ground and took approximately \SI{2}{\minute}\SI{30}{\second} for both ascending and descending with a maximum flight time of \SI{12}{\minute}. We tested the camera in photo configuration where we achieve a maximum framerate at full resolution of approximately 2.5fps. A sample image taken by the camera is presented in Fig.\ref{fig:view}. The drone was tested using two different flying modes. In the first one we perform a simple elevation scan, while CLASS was performing an azimuth scan. In the second one, we tested a raster scan while the telescope was kept fixed. Both flying modes have similar flight times and during both strategies we were compensating the elevation change of the drone with a gimbal movement to keep constant the relative elevation angle between the source and the telescope. 

Instead, in Fig.\ref{fig:adc} we presented the raw output of the multiplier as read by the diode and the ADC. As it is possible to notice there is a thermal drift starting when the drone starts to scan. This is likely due by the fact that the drone goes to the maximum velocity during the ascent increasing the air flow to cool the multiplier, while during the scanning the velocity is a bit lower. This result implies that in future flights, we need to use a different thermal paste to increase the contact between the multiplier and the aluminum frame. 

\section{FUTURE and CONCLUSION}
\label{sec:conclusion}

In this publication we show a new concept of a polarization calibration source for CMB experiments. This concept has been developed and tested on the field in its first iteration. The analysis is still ongoing but we can already say that the system performed well. We have some minor tweaks that we want to apply in the future version that will include a better thermal connection between the multiplier and the payload and some work to reduce the weight and consequently increase the flight time. We are also working to include new inclinometer with a faster sampling rate to improve the attitude reconstruction. Finally we are working to extend the frequency band and add other sources. 

\section{Acknowledgements}
Gabriele Coppi is supported by the European Research Council under the Marie Sk\l{}odowska Curie actions through the Individual European Fellowship No. 892174 PROTOCALC. RDP and FC are supported by FONDEF ID21I10236, HoverCal, and BASAL ACE210002. Giulia Conenna is supported by PRIN-MIUR 2017 COSMO.

\bibliography{main} 
\bibliographystyle{spiebib} 

\end{document}